\documentclass[modern]{aastex62}

\usepackage{graphicx,epsfig,natbib,color}
\usepackage{epstopdf}
\usepackage{subfigure}
\usepackage{threeparttable}
\usepackage{graphicx}
\usepackage{epsfig}
\usepackage{natbib}
\usepackage{multirow}

\shorttitle{shorttitle}
\shortauthors{Huang et al.}
\begin{document}

\title{\textbf{The Kinematic Evolution of Erupting Structures in Confined Solar Flares}}

\correspondingauthor{}
\email{xincheng@nju.edu.cn}

\author{Z. W. Huang}
\affil{School of Astronomy and Space Science, Nanjing University, Nanjing 210023, China\\}
\affil{Key Laboratory of Modern Astronomy and Astrophysics (Nanjing University), Ministry of Education, Nanjing 210093, China\\}
\author{X. Cheng}
\affil{School of Astronomy and Space Science, Nanjing University, Nanjing 210023, China\\}
\affil{Key Laboratory of Modern Astronomy and Astrophysics (Nanjing University), Ministry of Education, Nanjing 210093, China\\}
\affil{Institute of Physics and Astronomy, University of Potsdam, D-14476 Potsdam, Germany\\}
\author{M. D. Ding}
\affil{School of Astronomy and Space Science, Nanjing University, Nanjing 210023, China\\}
\affil{Key Laboratory of Modern Astronomy and Astrophysics (Nanjing University), Ministry of Education, Nanjing 210093, China\\}

\begin{abstract}
In this Letter, we study the kinematic properties of ascending hot blobs associated with confined flares. Taking advantage of high-cadence extreme-ultraviolet images provided by the Atmospheric Imaging Assembly on board the Solar Dynamics Observatory, we find that for the 26 events selected here, the hot blobs are first impulsively accelerated outward, but then quickly slow down to motionlessness. Their velocity evolution is basically synchronous with the temporal variation of the Geostationary Operational Environmental Satellite soft X-ray flux of the associated flares, except that the velocity peak precedes the soft X-ray peak by minutes. Moreover, the duration of the acceleration phase of the erupting blobs  is moderately correlated with that of the flare rise phase. For nine of the 26 cases, the erupting blobs even appear minutes prior to the onset of the associated flares. Our results show that a fraction of confined flares also involve the eruption of a magnetic flux rope, which sometimes is formed and heated prior to the flare onset. We suggest that the initiation and development of these confined flares are similar to that of eruptive ones, and the main difference may lie in the background field constraint, which is stronger for the former than for the latter.
\end{abstract}

\section{Introduction}
Solar flares, one of the most energetic of activities, occur in the atmosphere of the Sun and appear as sudden brightenings in various wavelengths. Solar flares are usually accompanied by coronal mass ejections (CMEs), which are violent ejections of solar plasma and magnetic field. \citet{2001ApJ...559..452Z} investigated the relationship between CMEs and flares and proposed three kinematic evolution phases of CMEs: the slow rise phase, the impulsive acceleration phase, and the propagation phase, which are closely related to the pre-flare phase, the main phase, and the decay phase of associated flares, respectively. Such a synchronization has also been found in studies with more advanced observations, such as \citet{Maricic_2007}, \citet{Bein_2012}, and \citet{Cheng_X_2020}. Moreover, \citet{Temmer_2008,Temmer_2010} and \citet{Qiu_2004} revealed a close correlation between the CME acceleration and the hard X-ray flux of flares. \citet{Berkebile-Stoiser_2012} even found a strong correlation between CME peak acceleration and spectral hardness of accelerated electrons. At present, it is widely believed that CMEs and flares are two manifestations of the same eruption process characterized by a violent disruption of the magnetic field in the corona, as is well explained by the two-dimensional CSHKP model \citep{Carmichael_1964,Sturrock_1966,Hirayama_1974,Kopp_Pneuman_1976} and its three-dimensional extension \citep{Aulanier_2012,Janvier_2013}. 

However, some statistical studies have revealed that not all flares are associated with CMEs (e.g., \citealt{Harrison_1995,Yashiro_2005}), even for a small number of extremely large X-class flares \citep{Green_2002,Thalmann_2015}. These CME-less events are generally named as confined flares for the sake of distinguishing them from CME-associated eruptive flares. It was found that confined flares tend to be temporally impulsive \citep{Kahler_1989} and spatially compact. In the era before high-resolution observations, confined flares were also considered to be compact flares \citep{Svestka_1986}, which could be interpreted in the context of the flux emergence model \citep{Heyvaerts_1977_APJ,Shibata_1992}. This model is actually an extension of the reconnection model for X-ray jets, where the compact flare is a consequence of the reconnection between the newly emerging flux with the background flux. In addition, confined flares may also stem from the interaction of two magnetic loops as found by \citet{Su_Yang_2013}, in which reconnection inflows and outflows were clearly observed.

Recently, an important finding is that some confined flares can be caused by failed filament/prominence eruptions. \citet{Ji_2003} first reported the observation of a failed filament eruption on 2002 May 27. \citet{Torok_2005} simulated the same event numerically and suggested that the failure of the erupting flux rope is due to its unsatisfaction of torus instability, i.e., the background field declines with height too slowly (e.g., \citealt{Liu_Y_2008, Cheng_x_2011}), or sometimes the decay index shows a saddle-like distribution (e.g., \citealt{Guo_Y_2010}). Moreover, some studies disclosed that confined (eruptive) flares tend to occur in the center (edge) of the active regions (e.g., \citealt{Wang_2007,Cheng_x_2011}). \citet{Baumgartner_2018} also found that the confinement of a flare can be caused by the fast change of the orientation of the polarity inversion line with height. On the other hand, a laboratory experiment revealed that a torus-unstable flux rope can also fail if the twist number of the flux rope is too low (i.e., the edge safety factor $q_a<0.8$; \citealt{Myers_2015}). Note that the rotation of the flux rope axis (e.g., \citealt{Hassanin_2016, Zhou_ZJ_2019}), the coronal helicity \citep{Nindos_2004}, the total unsigned magnetic flux \citep{Li_Ting_2020}, and the external reconnection between the flux rope and overlying loops \citep{Yang_SH_2019} were also suggested as possible factors determining whether a flare is eruptive or confined.



In addition to failed filament/prominence eruptions, erupting high-temperature structures, which we call hereafter hot blobs, can also lead to confined flares (e.g., \citealt{Cheng_x_2014, Song_2014}). Hot blobs have been discovered in the past decades using the two new high-temperature passbands (131 and 94 {\AA}) of the Atmospheric Imaging Assembly (AIA; \citealt{2012SoPh..275...17L}) on board the Solar Dynamics Observatory (SDO; \citealt{Pesnell_2012}). In this Letter, we select 26 confined flares and focus on the early kinematic evolution of the erupting hot blobs associated with the flares. In Section \ref{s2} we describe the data and event selection. In Section \ref{s3} we present the methods for analysis and the results, which is followed by a summary and discussions in Section \ref{s4}.


\section{Data and Event Selection}
\label{s2}
We mainly utilize the data from SDO/AIA, which images the solar atmosphere through 10 ultraviolet (UV) and extreme-ultraviolet (EUV) passbands, effectively covering a wide temperature range from 0.06 MK to 20 MK, with a cadence of 12 s and a spatial resolution of 1.2 arcsec \citep{2012SoPh..275...17L}. To observe the rising hot blobs, we select the 131 {\AA} passband, which is dominated by the emission of Fe VIII and Fe XXI lines centered at log[T]=5.7 and 7.0, respectively. The 171 {\AA} passband, dominated by the emission of Fe IX line centered at log[T]=5.8, is selected for observing the overlying loops of the rising hot blobs. Meanwhile, the 304 {\AA} passband, dominated by the emission of He II line centered at log[T]=4.7, is used to check the appearance of associated filaments. The soft X-ray (SXR) 1–8 {\AA} flux of flares is provided by Geostationary Operational Environmental Satellite (GOES). 

We select events associated with all confined flares in the period from 2010 November to 2014 December by the following criteria: (1) they should be close to the solar limb as observed from the AIA perspective with the heliocentric angles larger than 60 degrees, which ensures that the projection effect is not prominent; (2) the flare magnitude is required to be larger than C5.0, for the sake of ensuring a clear definition of the flare brightening area and the flare evolution phases based on the SXR flux; (3) they involve moving hot structures that appear as a roughly round blob with a clear front as observed from an edge-on view. Finally, we select 26 confined flare events to perform the kinematic analyses of the erupting blobs. The absence of associated CMEs is confirmed through inspecting images from both AIA and Large Angle and Spectrometric Coronagraph coronagraphs \citep{Brueckner_1995}. The parameters of all 26 events are listed in Table \ref{b1}, where the last two events are from  \citet{Cheng_x_2014} and \citet{Cheng_x_2018}. Events in the period of 2010--2012 can also be observed as disk events by Solar Terrestrial Relations Observatory (STEREO), enabling us to inspect the morphology of the corresponding flare ribbons.

\begin{figure}[htbp]
     \centering
     \includegraphics[width=2.9in]{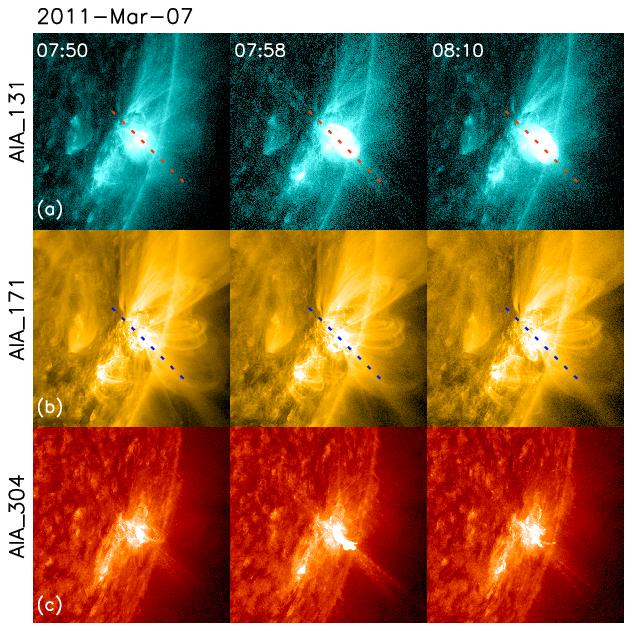}
     \includegraphics[width=2.9in]{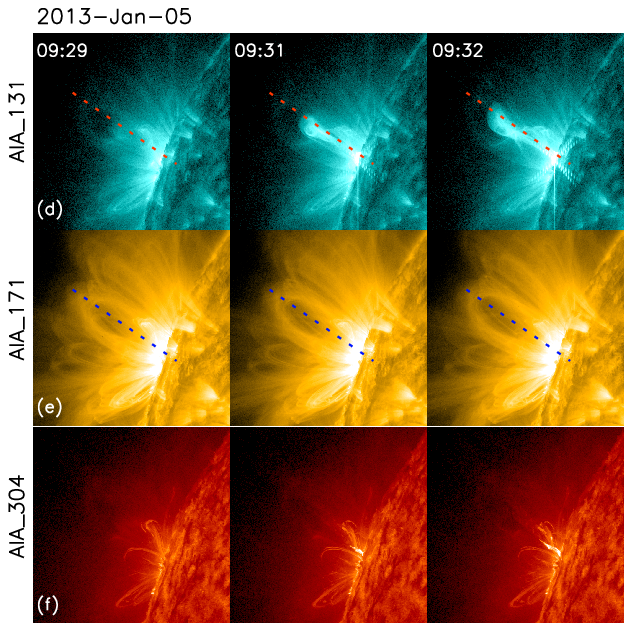}
     \caption{(a)-(c) SDO/AIA 131 {\AA}, 171 {\AA}, and 304 {\AA} images showing the evolution of the 2011 March 7 confined flare. (d)-(f) Same as (a)-(c) but for the 2013 January 5 confined flare. The dotted lines show the direction that we choose for constructing slice-time plots. \\
     (Animations of this figure are available in the electronic version.)}
     \label{f1}
\end{figure}

In Figure \ref{f1}, we present two typical examples of the confined flares that we select. The first event occurred on 2011 March 7 and was located at S20W78 (NOAA 11165). In Figure \ref{f1}(a), one can see a fast-rising blob in the AIA 131 {\AA} passband, which first appeared at around 07:45 UT and then slowly moved up. From 07:50 UT, the rising blob started to be rapidly accelerated until it was constrained by the overlying field at around 07:55 UT. Similar to the events in previous studies (e.g., \citealt{2011ApJ...732L..25C, Song_2014}), the blob was only visible in the AIA 131 and 94 {\AA} passbands, indicating its high-temperature property. The GOES data show that the associated flare started at 07:49 UT, about 4 minutes later than the first appearance of the blob, and peaked at 07:54 UT. In the 171 {\AA} passband (Figure \ref{f1}(b)), though overlying coronal loops above the rising blob can be clearly observed, no obvious lifting motion was found. In the 304 {\AA} passband (Figure \ref{f1}(c)), a filament eruption was also seen to first rise up and then fell back to the surface. However, the filament was much smaller than the hot blob. \citet{Cheng_x_2014} interpreted that the hot blob represents a flux rope system and that the filament only corresponds to the cool plasma collected in the dips of the flux rope.

The second confined flare occurred on 2013 January 5 and was located at N20E88 (NOAA 11652). It started at 09:28 UT and peaked at 09:31 UT. Different from the first event, no ascending structure in the AIA 131 {\AA} passband at the beginning of the flare was found until about 1 minute later (09:29 UT). After a period of uplift motion, the hot blob stopped at around 09:32 UT and failed to escape from the corona. In the 171 {\AA} passband (Figure \ref{f1}(e)), one can clearly see an expansion of the background coronal loop. \citet{Song_2014} also studied the kinematic evolution of this event and found that the velocity of the rising blob increased rapidly before reaching the peak value, with an extremely large acceleration ($\sim$7000 m s$^{-2}$) at the beginning of the eruption.


\begin{figure}
     \centering
     \includegraphics[width=6in]{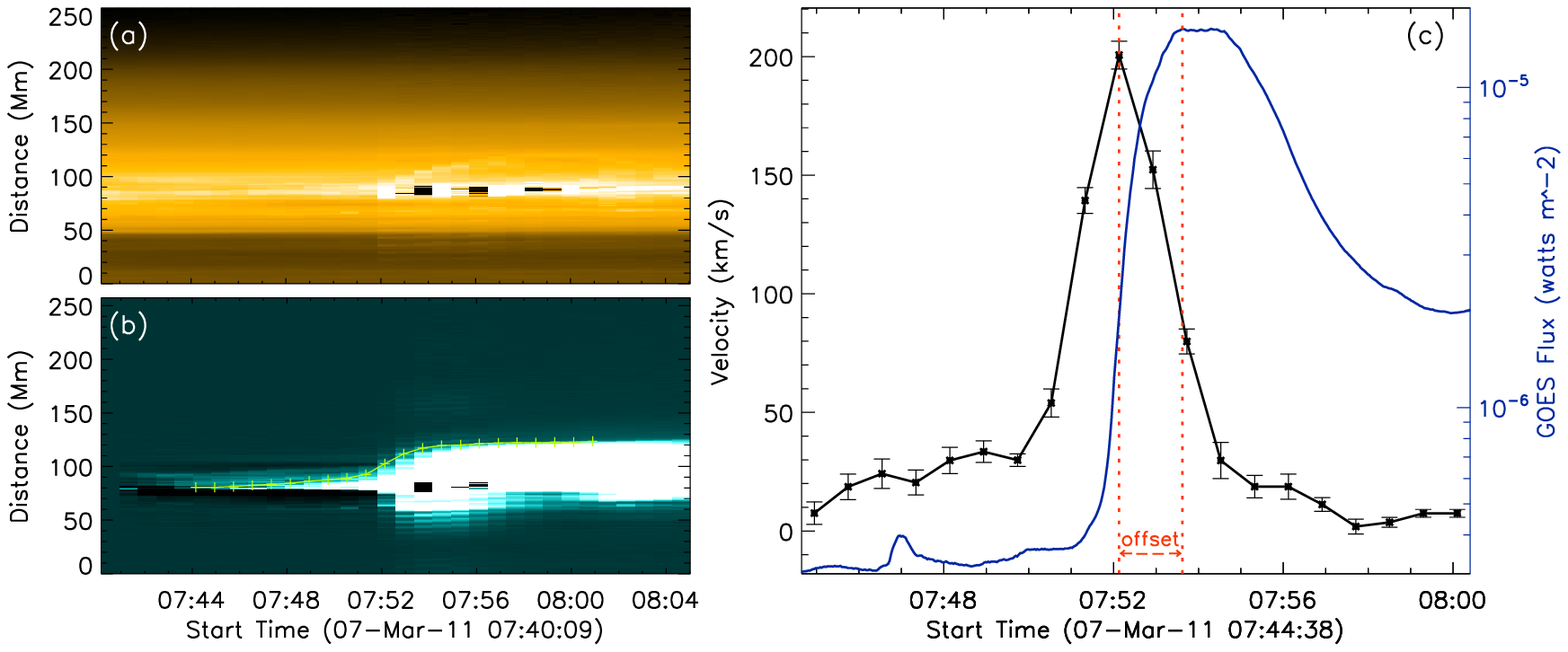}
     \includegraphics[width=6in]{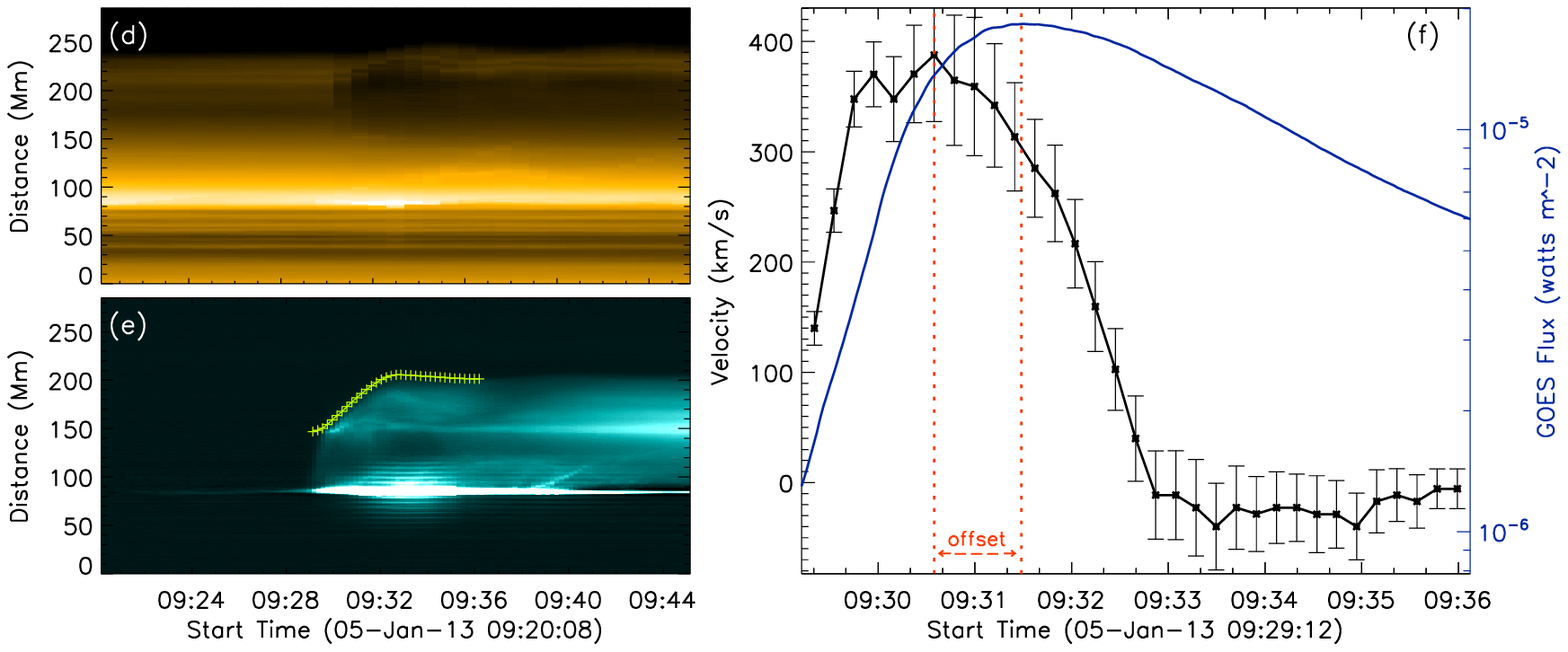}
     \caption{(a) and (b) Time-slice plots of the AIA 171 {\AA} and 131 {\AA} images showing the evolution of the overlying coronal loops and the hot blob during the 2011 March 7 confined flare. The green pluses indicate the measured distances for the top of the blob. (c) Time profile of the velocity of the blob (black), along with the time profile of GOES SXR 1–8 {\AA} flux. The peak time of the velocity and that of the SXR flux are marked by two orange dotted lines. (d)-(f) Same as (a)-(c) but for the 2013 January 5 confined flare.}
     \label{f2}
\end{figure}

\section{Results}
\label{s3}
\subsection{Early Kinematic Evolution}
In order to analyze the early kinematic evolution of the rising blobs and corresponding overlying loops, we first measure their time-varying heights based on the AIA 171 {\AA} and 131 {\AA} images. In Figure \ref{f2}, we show the time-slice plots for two examples. The location of the slice is selected to follow the eruption direction of the hot blob as shown in Figure \ref{f1}. Figure \ref{f2}(a) and \ref{f2}(d) show the time-slice plots at the 171 {\AA} passband. For the 2011 March 7 event, the upward motion of the overlying coronal loop was hardly detected, but the overlying loops became brighter with time, which is probably due to the compression of the rising blob. Note that part of the hot blob structure can also be observed in the time-slice plot at the 171 {\AA} passband, starting at around 07:50-07:52 UT (Figure \ref{f2}(a)). For the 2013 January 5 event, the ascending motion of the overlying loops was very clear, especially during 09:30-09:32 UT. Meanwhile, in the time-slice plots at the 131 {\AA} passband (Figure \ref{f2}(b) and \ref{f2}(e)), an evident rising blob was observed. Using the time-slice plots, we measure the heights of the tops of the ascending blobs at a time interval of about 48 s, which is sufficient to uncover their detailed evolution. Note that, for the flares with relatively short durations (7 of 26 events), we increase the cadence to 12 s.

Based on the height-time data, we calculate the velocities using a numerical differentiation method with three-point Lagrangian interpolation. The temporal evolutions of the velocity are shown in the right panels of Figure \ref{f2}. The errors of velocities are estimated from the standard deviation from multiple height measurements. For all events, we find that the rising blob presented an impulsive acceleration phase followed by a quick decrease phase before the velocity reached zero. For a fraction of events (9 of 26), like the 2011 March 7 event, we also observe a slow rise phase preceding the impulsive acceleration phase. For the remaining ones (17 of 26), like the 2013 January 5 event, the blob already had a considerable velocity or acceleration when it was first visible, without a detectable slow rise phase. We also compare the velocity evolution of the 2013 January 5 flare with the result of \citet{Song_2014} and find that they are consistent.

We further compare the kinematic evolution of the erupting blobs with the SXR emission of the associated flares. A very interesting finding is that the velocity evolution is basically synchronous with the temporal variation of the SXR flux, which is a typical feature for eruptive flares, except that for these confined flares, the SXR peak is delayed by minutes relative to the velocity peak. A careful inspection of the kinematic evolution of the erupting blobs for all confined flares studied here shows that this is a common feature. Moreover, for the confined events in the period of 2010-2012, we find that the flare morphologies also present two ribbons similar to eruptive flares by inspecting STEREO/Extreme Ultraviolet Imager (EUVI) images.




\subsection{Statistical Results}
We further study the offset in the peak times and that in the start times. The peak time offset is defined as the difference ($O_p=t_{fp}-t_{vp}$) between the peak time of hot blob velocity ($t_{vp}$) and that of the flare SXR emission ($t_{fp}$). The start time offset is defined as the difference ($O_s=t_{fs}-t_{vs}$) between the eruption onset time ($t_{vs}$) and the flare start time ($t_{fs}$). The eruption onset time is defined as the instant when the blob is first observed.

\begin{figure}
     \centering
     \includegraphics[width=6in]{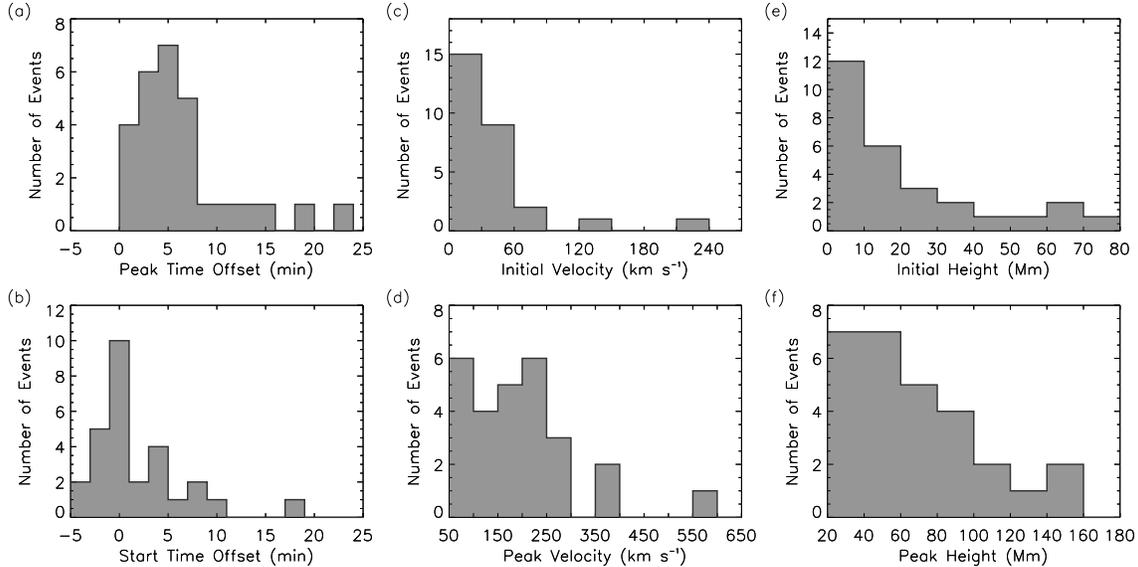}
     \caption{Distributions of the offset in peak times (a) and the offset in start times (b) for all the 26 events in the present study plus 2 events in the previous studies. (c) and (d) Initial and peak velocities of the hot blobs. (e) and (f) Initial and peak heights of the hot blobs.}
     \label{f3}
\end{figure}

The distributions of $O_p$ and $O_s$ for all the events are displayed in Figure \ref{f3}(a) and \ref{f3}(b). As shown in Figure \ref{f3}(a) and Table \ref{b1}, the peak time of the hot blob velocity obviously precedes that of the flare SXR flux for 22 of 26 confined flares, with an offset ranging from 2 to 22 minutes (larger than the measurement error of about 1 minute). For the other 4 cases, we find a weak offset of less than 2 minutes, which is still believable by using the data with a higher cadence of 12 s. In fact, the 2011 September 12 event in \citet{Cheng_x_2014} and the 2014 December 24 event in \citet{Cheng_x_2018} also show a similar offset in peak times. Therefore, such an offset in peak times could be a common feature for all confined flares. Note that, for the 2014 December 24 event, the erupting structure was a filament \citep{Cheng_x_2018}. 


Different from the offset in peak times, the offset in start times, $O_s$, does not show a single sign. It is less than -2 minutes for 7 events, greater than 2 minutes for 10 events, and marginally around zero (in the range of -2 to 2 minutes) for the remaining ones, among the 28 events in total, as shown in Figure \ref{f3}(b) and Table \ref{b1}. For the events involving the rising blobs that appeared obviously earlier than the start time of SXR flux, it is likely that the hot blobs are formed and heated before the flare, especially for the 4 events with an offset in start times of over 5 minutes. As a comparison, \citet{Nindos_2020} found that for most eruptive flares, hot flux ropes can be observed before the flare occurrence. On the other hand, for events with a significant negative offset in start time, it is likely that the erupting blob is newly formed by the flare reconnection and thus it first appears after the start of the flare. 

In Figure \ref{f3}(c) and \ref{f3}(d), we show the distribution of initial and maximum velocities of the hot blobs for each event. One can see that the initial velocities for most events are relatively low, except for the 2013 January 5 and 2014 March 12 events, whose initial velocities exceed 100 km s$^{-1}$. However, for these two events, the start times of the eruptions are significantly later than that of the flares; thus, it is possible that in the very early phase of the event, the rising blob is too cool to be detectable. The peak velocities of the hot blobs in these confined flares lie in the range of 60--500 km s$^{-1}$, much smaller than that of 100--1600 km s$^{-1}$ for eruptive events \citep{Zhang_2006}. This is simply due to the fact that the acceleration phase of the erupting structure in confined flares is much shorter than that in eruptive flares. Nevertheless, the velocity of the rising blobs is still much larger than that of the rising post-flare loops, which ranges from several km s$^{-1}$ (e.g., \citealt{Cheng_X_2010_re}) to several tens km s$^{-1}$ \citep{Veronig_2006_AnA}. In addition, as shown in Figure \ref{f3}(e), the initial heights of the top of hot blobs in all the events are within 70 Mm with an average of 20 Mm, which is comparable with that of the erupting filaments (e.g., \citealt{Filippov_2000}). Note that the projection effect has been corrected by assuming radial motions of the blobs.

\begin{figure}
     \centering
     \includegraphics[width=4in]{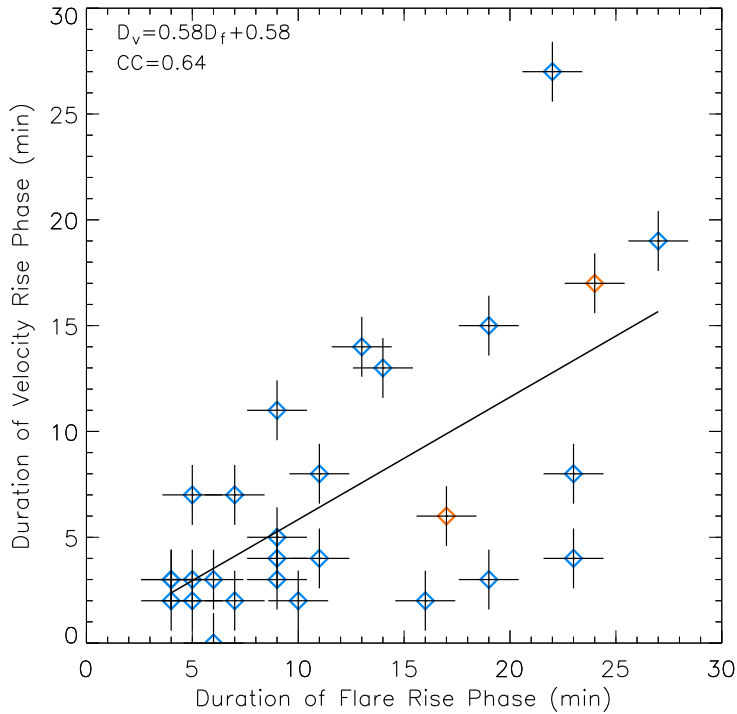}
     \caption{Scatter plot of the duration of the SXR rise phase versus that of the velocity rise phase for all the 26 events in this study (blue) and 2 events from previous studies (orange), with a mean error of 1.4 minutes. The oblique line shows a linear fit to all data points.}
     \label{f4}
\end{figure}

Moreover, we also calculate the duration of the velocity rise phase ($D_v=t_{vp}-t_{vs}$) and that of the flare rise phase ($D_f=t_{fp}-t_{fs}$) and study their relationship. All parameters mentioned above are shown in Table \ref{b1}. A scatter diagram of the two durations is shown in Figure \ref{f4}. One can clearly see that the duration of the flare rise phase is moderately related to that of the velocity rise phase, with a correlation coefficient of 0.64. A linear fitting of the data points yields $D_v=0.58D_f-0.58$. It shows that the duration of the acceleration phase of the hot blob is systematically shorter than that of the flare rise phase.

\section{Summary and Discussion}
\label{s4}
In this Letter, we study the property of the erupting structures during 26 confined flares close to the solar limb, mainly focusing on their early kinematic evolution. The erupting structures appeared as hot blobs and experienced a two-phase evolution, i.e., an impulsive acceleration phase followed by a fast deceleration one. Moreover, the velocity evolution of the hot blobs is found to be roughly synchronized with the SXR light curve of the associated confined flares. These features are very similar to the kinematic characteristics of the erupting flux ropes during eruptive flares as found by \citet{Cheng_X_2020}.

However, we also find an offset between the peak time of the velocity and that of the flare SXR flux, which could be due to the strong constraint of the overlying field the rising hot blob. We tend to interpret the confined flares involving the erupting hot blobs with the standard CME/flare model (i.e. CSHKP model) instead of loop-loop reconnection \citep{Su_Yang_2013}, flux emergence model \citep{Veronig_2015_SoPh}, or single-loop instability \citep{Sakai_1996_SSRv}, in the latter three of which the erupting hot blob is not expected. For eruptive events, as interpreted by the CSHKP model, the eruption of the CME flux rope forms a long current sheet connecting the erupting flux rope and the flare loop. Once magnetic reconnection in the current sheet is switched on, more and more new flux is added to the flux rope and facilitates its eruption. The flux rope eruption in turn enhances the reconnection and thus the flare emission (e.g. \citealt{Vrsnak_2016_AN,Veronig_2018_ApJ}). This forms a positive feedback process between the eruption of the CME flux rope and magnetic reconnection giving rise to the flare emission. Thus, the evolution of the CME is almost synchronized with that of the associated flare, as discussed in \citet{Cheng_X_2020}. Nevertheless, for confined flares, the eruption is suppressed by the overlying field, so that the flux rope may experience an evolution of deceleration following acceleration. However, shortly after the velocity reaches its peak, the hot blob is decelerated but still ascends and stretches the background coronal loops, so that the reconnection continues and the flare emission is still being enhanced. Only when the hot blob reaches its maximum height and stops rising, does the reconnection stop. This results in an offset between the velocity peak time and the flare peak time. Such an offset could be a reason why the correlation between the time profile of the velocity and that of the flare emission variation is stronger for eruptive flares \citep{Zhang_2006} than for confined flares. Based on the statistical result of \citet{Nindos_2015}, who found that 11 out of 64 M-class and X-class confined flares have a signature of hot flux ropes, the scenario that we propose here can at least apply to about 17\% of all major confined flares. Note that, for eruptive flares, \citet{Reeves_2006_ApJ} pointed out that the different reconnection rate and background magnetic field strength may also cause the offset found here.


We conjecture that the initiation and development of confined flares with hot blobs, at least for the events that we study, may be very similar to that of eruptive flares. The main difference is that the overlying fields over confined flares decay more slowly, as has been found in some previous studies focusing on the comparison between confined and eruptive flares (e.g., \citealt{Liu_Y_2008, Cheng_x_2011,Sun_XD_2015,Wang_d_2017,Baumgartner_2018}). It is worth mentioning that the previous statistical studies were mostly based on the assumption that each eruption involves a flux rope. However, such an assumption has not been fully proved, in particular for confined flares. Here, our detection of erupting hot blobs provides strong evidence for the existence of flux ropes in confined flares, considering their similar kinematic evolutions and flare morphologies to eruptive flares. Recently, \citet{Hernandez_2019} studied a limb confined flare and reported a hot cusp structure likely linked with kinked flare loops, which may also indicate the existence of a flux rope.

\acknowledgements  

We thank the referee for their helpful comments and suggestions. We also thank the team of SDO/AIA and STEREO/SECCHI for providing the valuable data. AIA data are courtesy of NASA/SDO, which is a mission of NASA’s Living With a Star Program. STEREO is a mission in NASA’s Solar Terrestrial Probes program. Z.W.H., X.C., and M.D.D. are funded by NSFC grants 11722325, 11733003, 11790303, 11790300, 11961131002, Jiangsu NSF grants BK20170011, and “Dengfeng B” program of Nanjing University. 

\newpage
\begin{table}[!htbp]
     \caption{Parameters of 26 confined flares}
     \centering
     \begin{tabular}{l*{10}{c}}
     \hline
     \hline
     \multirow{2}*{No.} & \multirow{2}*{Date} & \multirow{2}*{Flare Class} & \multicolumn{2}{c}{Duration$^{a}$ (min)} &\multicolumn{2}{c}{Offset (min)}& \multicolumn{2}{c}{Velocity (km s$^{-1}$)}&\multicolumn{2}{c}{Height (Mm)}\\
     \cline{4-11}
       & & & SXR & Velocity & Start & Peak & Start & Peak & Start & Peak\\
     \hline 
     1.& 2010 Nov 5 & M1.0 & 9 & 3 & -3 & 3 & 17 & 105 & 70 & 120\\
     2. & 2010 Nov 6 & M5.4 & 9 & 4 & 0 & 5 & 83 & 209 & 25 & 159\\
     3. & 2011 Mar 6 & C7.5 & 9 & 11 & 5 & 5 & 18 & 217 & 4 & 47\\
     4. & 2011 Mar 7 & C5.0 & 19 & 15 & 3 & 7 & 11 & 53 & 15 & 37\\
     5. & 2011 Mar 7 & M1.5 & 5 & 7 & 4 & 2 & 8 & 201 & 17 & 57\\
     6. & 2011 Mar 8 & M5.3 & 9 & 5 & 0 & 4 & 50 & 141 & 4 & 55\\
     7. & 2011 Mar 10 & M1.1 & 7 & 7 & 2 & 2 & 11 & 88 & 8 & 39\\
     8. & 2011 Mar 15 & M1.0 & 4 & 3 & -1 & 0 & 42 & 287 & 11 & 27\\
     9. & 2011 May 1 & C6.5 & 7 & 2 & -1 & 4 & 28 & 171 & 14 & 29\\
     10. & 2011 May 27 & C5.6 & 5 & 3 & 0 & 2 & 34 & 133 & 2 & 29\\
     11.& 2011 May 28 & C8.3 & 27 & 19 & 0 & 6 & 11 & 95 & 16 & 69\\
     12.& 2011 Jun 14 & M1.3 & 11 & 4 & -2 & 5 & 52 & 210 & 1 & 76\\
     13.& 2011 Sep 4 & C5.8 & 13 & 14 & 8 & 7 & 7 & 27 & 49 & 66\\
     14.& 2011 Sep 11 & C6.6 & 10 & 2 & -1 & 7 & 62 & 250 & 24 & 80\\
     15.& 2011 Sep 21 & M1.8 & 19 & 3 & -1 & 15 & 55 & 195 & 2 & 38\\
     16.& 2011 Sep 24 & M1.7 & 23 & 8 & 7 & 22 & 6 & 72 & 35 & 79\\
     17.& 2012 Mar 23 & M1.0 & 6 & 3 & -2 & 1 & 19 & 212 & 0 & 54\\
     18.& 2012 Apr 19 & C7.0 & 11 & 8 & 4 & 7 & 9  & 98 & 0 & 45\\
     19.& 2012 Oct 10 & C5.1 & 4 & 3 & 1 & 2 & 37 & 164 & 4 & 22\\
     20.& 2012 Oct 21 & C7.8 & 14 & 13 & 9 & 10 & 15 & 261 & 4 & 45\\
     21.& 2012 Oct 23 & X1.8 & 4 & 2 & -1 & 1 & 12 & 576 & 8 & 105\\
     22.& 2012 Nov 30 & C5.4 & 23 & 4 & -1 & 18 & 33 & 107 & 6 & 70\\
     23.& 2013 Jan 5 & M1.7 & 5 & 2 & -3 & 0 & 140 & 388 & 62 & 117\\
     24.& 2013 Dec 20 & M1.6 & 22 & 27 & 18 & 13 & 14 & 89 & 56 & 152\\
     25.& 2014 Mar 12 & M9.3 & 6 & 0 & -4 & 2 & 230 & 390 & 32 & 91\\
     26.& 2014 Mar 13 & M1.2 & 16 & 2 & -5 & 8 & 47 & 152 & 29 & 55\\
     27.& 2011 Sep 12 & C9.9 & 24 & 17 & -2 & 5 & 41 & 236 & 64 & 99\\
     28.& 2014 Dec 24 & C3.7 & 17 & 6 & 3 & 4 & 12 & 195 & 19 & 96\\
     \hline
     \end{tabular} 
     \label{b1}
\end{table}

\acknowledgements  
\newpage


\newpage
\pagenumbering{Roman}


\clearpage

\end{document}